\documentclass[twocolumn]{aastex631}

\usepackage{graphicx} 
\date{June 2024\vspace{-1cm}}
\usepackage{graphicx}
\usepackage{color}
\usepackage{amsmath}
\usepackage[english]{babel}
\usepackage{times}
\usepackage[]{natbib}
\begin{document}

\title{The Excess of JWST Bright Galaxies: a Possible Origin in the Ground State of Dynamical Dark Energy in the light of DESI 2024 Data
} 
\author{N. Menci$^1$, A. A. Sen$^2$, M. Castellano$^1$}
\affil{$^1$ INAF - Osservatorio Astronomico di Roma, via Frascati 33, I-00078 Monte Porzio, Italy}
\affil{$^2$Centre For Theoretical Physics, Jamia Millia Islamia, New Delhi, 110025, India.}
\begin{abstract} 
Recent observations by JWST yield a large abundance of luminous galaxies at $z\gtrsim 10$ compared to that expected in the $\Lambda$CDM scenario based on extrapolations of the star formation efficiency measured at lower redshifts. While several astrophysical processes can be responsible for such observations, here we explore to what extent such an effect can be rooted in the assumed Dark Energy (DE) sector of the current cosmological model. This is motivated by 
recent results from different cosmological probes combined with the  last data release of the Dark Energy Spectroscopic Instrument (DESI), which indicate a tension in the DE sector of the concordance $\Lambda$CDM model. 
We have considered the effect of assuming a DE characterized by a negative $\Lambda$ as the ground state of a quintessence field on the galaxy luminosity function (LF) at high redshifts. We find that such models naturally affect the galaxy UV luminosities in the redshift range $10\lesssim z\lesssim 15$ needed to match the JWST observations, and with the value of $\Omega_{\Lambda}=[-0.6,-0.3]$ remarkably consistent with that required by independent  cosmological probes. A sharp prediction of such models is the steep decline of the abundance of bright galaxies in the redshift range $15\lesssim z\lesssim 16$. 
\end{abstract}


\section{Introduction}

The large number  density of UV-bright galaxies measured by JWST at redshift $z\gtrsim 9$ appreciably exceeds the expectations of simulations and models based  on the standard $\Lambda$CDM cosmology and on the extension of the star formation efficiency measured at lower redshifts \citep[e.g.][]{Castellano2022b,Castellano2023a,Finkelstein2023a,Finkelstein2023b}. Many physical interpretations of such a tension have  been discussed in the literature. Among these, feedback-free regime at high $z$ \citep[e.g.][]{Dekel2023}, top-heavy stellar initial mass function \citep[e.g.,][]{Trinca2024},  negligible dust attenuation \citep[e.g.,][]{Ferrara2023}, stochastic star formation \citep[][]{Mason2022,Kravtsov2024} constitute viable astrophysical explanations. 

While all the above solutions concern the complex physics relating star formation to the evolution of dark matter (DM) haloes at high redshifts, 
there are hints that the problem may be rooted in the cosmological framework. E.g., the large number density of massive galaxies already in place at such large redshifts \citep[e.g.,][]{Labbe2023,Xiao2023,Casey2024} seems to be on the verge of challenging the $\Lambda$CDM models even assuming a maximal efficiency $\epsilon=1$ for the conversion of baryons into stars at earlier epochs \citep[e.g.,][]{Menci2022,BoylanKolchin2022,Lovell2023}. 
Although the stellar mass estimates of these galaxies are uncertain and even their identification is subject to debates \citep{Endsley2023,Kocevski2023,Chworowsky2024}, confirmation of such results would imply the need for a revision of the cosmological model. In particular, the abundance of such galaxies might be easily accounted for by assuming a dynamical (i.e. time evolving) DE equation of state parameter $w=p/\rho$ is characterised by {\it phantom} behaviour $w<-1$ at early epochs, and a larger value $w>-1$ at present times \citep{Menci2022}.

What makes such an explanation particularly interesting is that recent measurements from independent cosmological probes concur in indicating that the problem may be rooted in the DE sector of the current cosmological scenario. In fact, recent breakthroughs from baryonic acoustic oscillations (BAO) measured by the DESI \citep{Adame2024} collaboration, when combined with the cosmic microwave background (CMB) observations by Planck and with the luminosity distance measurements through Type Ia Supernovae (SNIa) have shown deviations from the $\Lambda$CDM predictions (within a two-parameter DE model) estimated as 2.5 $\sigma$, 3.5 $\sigma$, and 3.9 $\sigma$ depending on the supernova dataset included in the compilation (respectively, PantheonPlus, Union3, and DESY5). Although these results depend on a particular kind of  parametrisation for the evolution of DE, the hint of phantom  crossing in the DE equation of state has also been inferred for different DE  paramatrisations \citep{lodha2024, giare2024} as well as with model independent analysis \citep{calderon2024}. 
The same phantom behaviour of DE has also been shown to constitute a possible explanation of the JWST observations of massive galaxies at $z\gtrsim 6$, \citep[see][]{Menci2022,Cortes2024}. 

While the above measurements strongly suggest that the expansion history of the Universe differs from that envisaged by the concordance $\Lambda$CDM cosmological model, a DE equations of state with $w<-1$ is extremely problematic for all the  scenarios in which 
the cosmic acceleration is traced back to the dynamics of a scalar field $\phi$. In all these scenarios  a scalar field $\phi$ rolling in a potential $V(\phi)$ would yield an equation of state-parameter $w=(\dot\phi^2-V(\phi))/(\dot\phi^2+V(\phi))$. While such scenarios are attractive because - for a proper form of the potential $V(\phi)$ - they would provide a natural way to achieve  a negative equation of state and an accelerated expansion (in analogy to the mechanism at the basis of cosmic inflation), achieving $w<-1$ would require a negative kinetic term  \citep[see][for a review]{Ludwick2017}.  However, an expansion history consistent with all the observations mentioned above can be achieved considering a field  whose potential features a negative minimum (Anti-de Sitter/AdS vacuum). In this case, the positive energy density $\rho_x>0$ of the evolving DE component on top of a negative cosmological constant (nCC) must yield a net positive value $\rho_x+\rho_{\Lambda}> 0 $ around the present time, so as to be consistent with the observed late-time acceleration. 

Such scenarios have been widely investigated in the literature. Besides the  solid theoretical motivation for the presence of a nCC (Antonini et al. 2023, Demirtas et al. 2022), such models  have been proved to  perform equally well as $\Lambda$CDM, or are even potentially statistically preferred, when confronted to a number of cosmological probes \citep[see, e.g.,][]{Calderon2021,Sen2022,Adil2023}. 
In addition, nCC models can help to reduce the well known  5$\sigma$ tension between early and late Universe inferences of the Hubble constant $H_0$ \citep[e.g.,][]{Riess2022,Planck2020} in the $\Lambda$CDM scenario \citep{Sen2022}. Finally, the nCC scenario provides a better match to the observed abundance of massive galaxies even assuming a quintessence (i.e. non phantom) DE equation of state \citep{Menci2024}.

In this context, here we show that nCC models are also characterized by a boost in the characteristic mass for collapse with respect to $\Lambda$CDM that provides a potential cosmological explanation for the observed standstill in the evolution of the bight end of the UV luminosity functions at $z> 9$. Although it is perfectly possible that one (or more) of the astrophysical processes described above can be responsible for such observations, it is intriguing that, while the  astrophysical processes need to be tuned to modify the galaxy $L/M$ ratios in the redshift range $z\approx 10-15$, nCC models provide a natural way to account for such a standstill in this redshift range, without the need for sharp changes in the physics of galaxy formation.

\section{Dark Energy Models with Negative Cosmological Constant}

A feature common to most scalar field DE models is the positivity of the the ground state of the field potential $V(\phi)$,  corresponding to a stable or meta-stable de Sitter (dS) vacuum. In the simplest model, 
 the scalar field $\phi$ is settled at this minimum, resulting into a positive cosmological constant that can drive the accelerated expansion of the Universe. A more general scenario is when the field $\phi$ is not settled at the minimum of $V(\phi)$ but rolls slowly over the potential and we get a dynamical dark energy, popularly termed as "quintessence". Unfortunately constructing such a quintessence field with de-Sitter ground state is extremely challenging in quantum gravity theories. In fact according to Swampland Conjecture, a dS ground state or at least a stable dS ground state can not appear in any reliable string theory construction, see \citet{Vafa}, \cite{Agrawal}. On the other hand a scale field rolling over a potential with negative minimum (also known as anti de-Sitter (AdS) minimum or ground state) is a common feature in string theory; one of the reasons being the famous anti-de Sitter/conformal field theory (AdS-CFT) correspondence \citep{Maldacena1998}
 as well as due to holography \citep{raam}. AdS ground state for scalar field potentials results in the presence of a negative cosmological constant, and using scalar fields with such potentials having AdS minimum is a perfectly viable model for quintessence. 
 
 Motivated by the above considerations, here we study some striking implications of such quintessence models with AdS vacua on galaxy formation. Instead of taking any particular potential for the scalar field (which would restrict us to a specific model), we parametrise the dynamical nature of its equation of state $w(a)$ using the most popular Chevallier-Polarski-Linder (CPL) parametrisation, see \citet{Chevallier2001} and \citet{Linder2003} :

\begin{equation}
w(a) = w_{0} + (1-a) w_{a},
\end{equation}

\noindent
where $a$ is the expansion factor and $w_{0}$ and $w_{a}$ are two arbitrary constants. Scalar field models with potentials having an AdS ground state are represented by a dynamical part $\rho_x$ having an equation of state $w$ given by eq. (1) together with a cosmological constant $\rho_{\Lambda}$ which is negative. Finally, the equation governing the expansion of the Universe in such a model is given by:

\begin{equation}
    \left[\frac{H(a)}{H_0}\right]^2 = \Omega_{m}a^{-3} + \Omega_{\Lambda} + \Omega_{x} f(a),  
\end{equation}

\noindent
where $ f(a) = a^{-3(1+w_{0} + w_{a})} exp[-3w_{a}(1-a)]$ and $\Omega_{m}+\Omega_{\Lambda}+\Omega_x=1$ due to spatial flatness. In the following we shall keep the value of the matter density parameter $\Omega_{m}=0.31$, with the normalization of the Hubble parameter $H_0=67$
 km s$^{-1}$ Mpc$^{-3}$, \citep[e.g.,][]{Planck2020}. 

Thus, the total density parameter of the DE sector is 
$\Omega_{DE}=\Omega_{\Lambda}+\Omega_x$.
 Although $\Lambda$ itself and therefore $\Omega_{\Lambda}$ can be negative, the total DE density and therefore $\Omega_{DE}$ have to be positive in order to be able to drive the observed cosmic acceleration at low redshifts and maintain agreement with cosmological observations. There have been number of studies in recent time that shows that such a DE model containing a negative cosmological constant is consistent with different cosmological observations. These include the consistency of this model with CMB (as observed by Planck-2018), BAO (as observed by SDSS) as well as SnIa measurements of Pantheon Sample \citep{Sen2022}. Subsequently DE models with negative $\Lambda$ has been confronted with Pantheon-Plus compilation of the SnIa observations \citep{Malekjani2023}, with CMB (Planck 2018)+BAO(SDSS)+Pantheon-Plus+SH0ES \citep{Adil2024}, with JWST photometric and spectroscopic observations of high redshift galaxies \citep{Menci2024} and more recently with DESI BAO measurements \citep{Wang2024}. Moreover the possible constraints on DE models with negative $\Lambda$ from near future SKA-mid observations has been also studied recently \citep{Dash2024}.

\section{Method}

To explore the impact of assuming negative values of $\Omega_{\Lambda}$ on galaxy formation, we consider  the evolution of the characteristic mass for the collapse of perturbations, defined as the mass $M_c(a)$ at which the 
 rms value of density perturbations equals the linear density threshold $\delta_c$ for collapse, i.e., 

 \begin{equation}
\sigma(M_c)\,D(a)=\delta_c
 \end{equation}
 where $\sigma^2(M)$ is the variance of the linear density field smoothed on the mass scale $M$, and $D(a)\equiv \delta (a)/\delta(1)$ is linear growth factor $D(a)$ accounting for the evolution of the linear  density field $\delta$.
 This provides the mass at which the exponential factor in any Press \& Schechter-like mass distributions of DM halos   ${dN/dM}= e^{-\delta_c^{2}/2\sigma^2(M)D^2(t))}$ begins to bend down the distribution. 
In fact, in terms of the characteristic mass the  \cite[][]{p&s} mass function can be written as 
$N(M)=\sqrt{(2/\pi)}A\,\rho\,\,M_c^{-2}\,(M/M_c)^{A-2}\,exp[-(M/M_c)^{2A}/2]$, 
where $A=(n_{eff}+3)/6$, and $n_{eff}$ is the effective spectral index of density perturbations at the mass scale $M$. 

 Assuming a CDM form for $\sigma(M)$ \citep{bardeen86}, we can derive the characteristic mass $M_c$ after eq. (1) for any nCC cosmology by computing the growth factor of density perturbations $D(a)$. This is obtained by numerically solving the equation governing the linear growth of density perturbations \citep[see][]{Adil2023}:
\begin{equation}
\delta''+\Bigg[{3\over a}+{E'(a)\over E(a)}\Bigg]\delta'={3\over 2}{\Omega_m\over a^5 E^2(a)}\delta
\end{equation}
where ' indicates a derivative with respect to the scale factor $a$, and $E(a) \equiv H(a)/H_0$ denotes the normalized expansion rate.

\vspace{0.1cm}
\hspace{-0.7cm}
\includegraphics[width=1\linewidth]{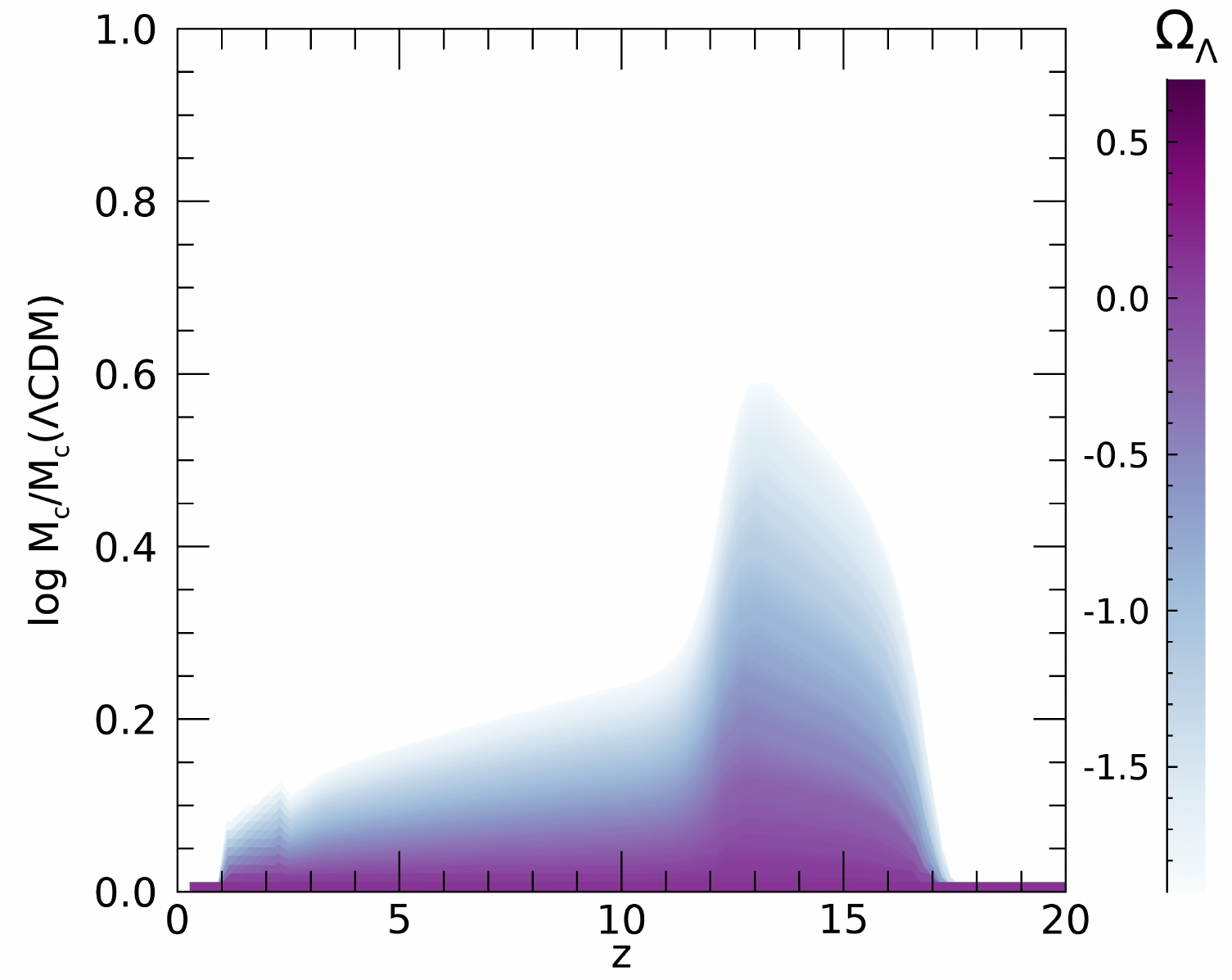}\newline
{\footnotesize Fig. 1 - The ratio of the characteristic mass $M_c(t)$ for different nCC models to that in the $\Lambda$CDM case, for our fiducial combination $w_0=-0.9$, 
$w_a=0.1$. The color code corresponds to different values of $\Omega_{\Lambda}$ as shown in the color bar.}
\vspace{0.2cm}

The evolution of the characteristic mass is shown in fig. 1 for  cosmological models with different value of the vacuum energy density parameter $\Omega_\Lambda$. The case with $\Omega_{\Lambda}=0.7$ corresponds to the standard $\Lambda$CDM cosmology. In all cases, 
we assume a fiducial combination $w_0=-0.9$, 
$w_a=0.1$. This is chosen as representative of a non-phantom, quintessence behaviour for the field responsible for the DE, which is also consistent with the most recent DESI data \citep[see][]{desicoll}. We  discuss below how assuming different combinations ($w_0, w_a$) affects our results. 

The striking feature in Fig. 1 is the boost in $M_c(z)$ compared to the $\Lambda$CDM case for $10\lesssim z\lesssim 15$. This coincides with the redshift range where current observations are showing an exceeding large abundance of bright galaxies compared to theoretical models and to extrapolations of the LF measured at lower redshisfts \citep[e.g.,][]{Finkelstein2023b}. The effect of such a boost on the DM mass function is illustrated in fig. 2, where we compare the evolution of the $\Lambda$CDM Press \& Schechter  mass function with the corresponding evolution in  three selected nCC models with $\Omega_{\Lambda}=-0.7$, 
$\Omega_{\Lambda}=-0.5$, and $\Omega_{\Lambda}=-0.2$ (the combination ($w_0$, $w_a$) is left fixed to our fiducial values). While at $z=7$ the effect of assuming nCC cosmologies is negligible, at redshift 
$z\gtrsim 10$ the DM halo mass function gets a significant boost, whose magnitude  grows with progressively smaller (more negative) values of $\Omega_{\Lambda}$. 

\vspace{0.1cm}
\hspace{0.cm}
\includegraphics[width=0.9\linewidth]{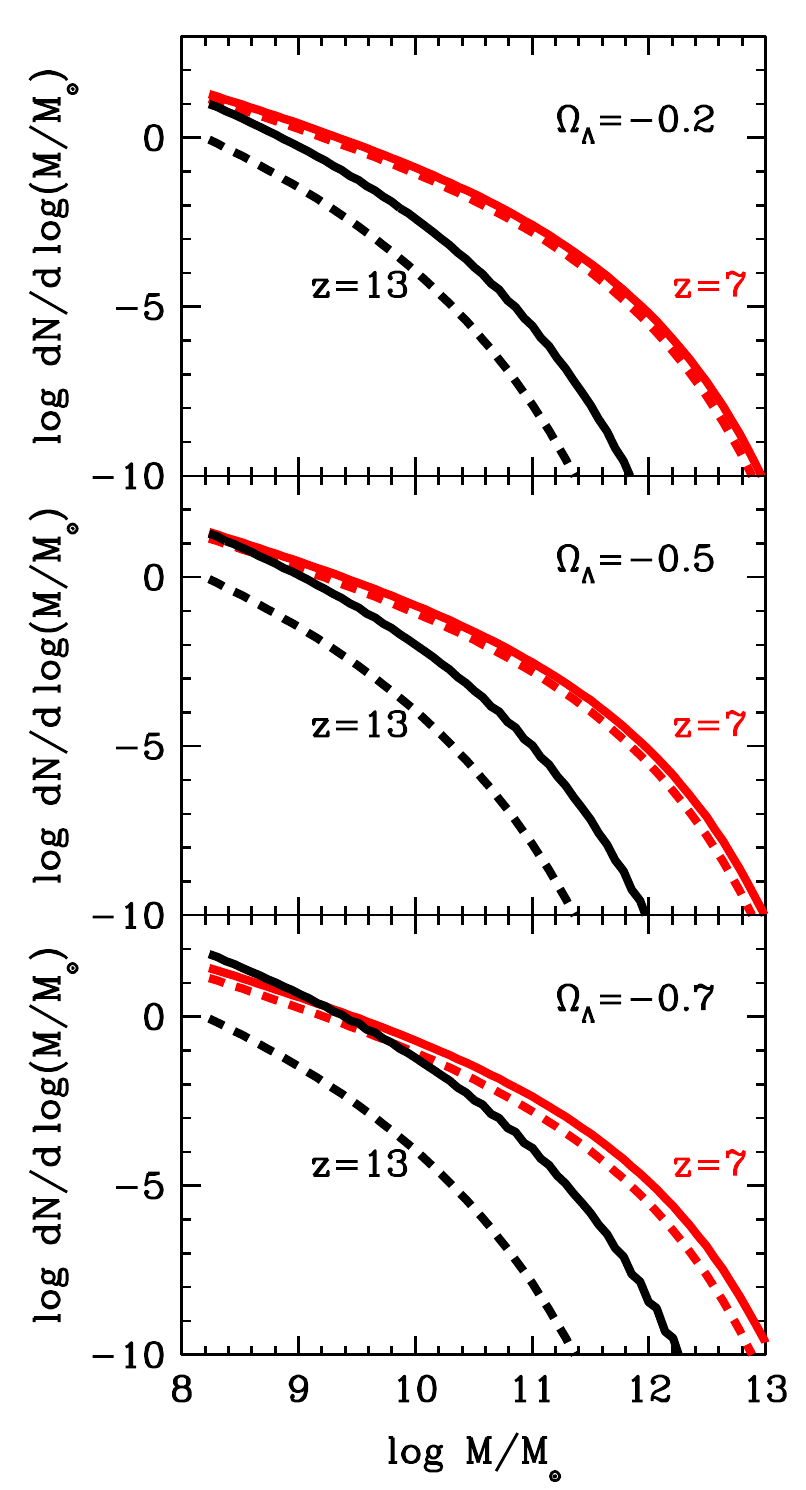}\newline
{\footnotesize Fig. 2 - The DM halo mass function corresponding at redshift $z=7$ (red lines) and $z=13$ (black lines) in nCC models with different, negative values of $\Omega_{\Lambda}$ aree shown as continuous lines. For reference, we  also show as dashed  lines the halo mass function corresponding to the standard $\Lambda$CMD cosmology.}
\vspace{0.1cm}

The physical origin of such a boost is rooted in the dependence of the growth rate of perturbations $D(a)$ on the  expansion factor $H(a)$. In fact, a faster expansion inhibits the growth of density perturbations, due to the larger dilution of density perturbations (see, e.g., eq. 2). Thus, the dependence of $H(a)$ on the assumed cosmology (see eq. 1)  critically affects the growth factor. At very early times $a\rightarrow 0$  the strong dependence of the term related to matter density $\Omega_m\,a^{-3}$ dominates over all other terms, so that $H(a)$ - and hence  a characteristic mass $M_c(a)$ - is almost independent on the other cosmological parameters $\Omega_{\Lambda}$ and $\Omega_{\phi}$ (as shown by the converging behaviour of $M_c(z)$ for $z\gtrsim 14$ in fig. 1) . However, at lower redshifts, the larger values of $a$ allow the value  of $\Omega_{\Lambda}$ to appreciably affect $H(a)$; eq. (1) shows that for decreasing values of $\Omega_{\Lambda}$ (and particularly for negative values)  a smaller expansion rate $H(a)$ is obtained, resulting into larger growth factors. This explains the increase of $M_c(a)$ compared to the $\Lambda$CDM case corresponding to the bump in fig. 1. Finally,  for $a\rightarrow 1$ the evolution of the expansion rate in eq. (1) - and hence the growth factor $D(a)$, and the characteristic mass $M_c(a)$ - reduces exactly to the $\Lambda$CDM case (since $\Omega_{x}+\Omega_{\Lambda}=1-\Omega_m\approx 0.7$, see Sect. 2).

To test whether such a boost is quantitatively able to account for the observations, we compute the LFs corresponding to nCC cosmologies. We start from the mass function $dN(M)/dM$ of DM haloes in nCC cosmologies,  computed for different values of $\Omega_{\Lambda}$ following the lines in \citet{Menci2022}. The DM mass $M$ is then related to the star formation rate of galaxies  $\dot m_* = \epsilon (M)\,f_b\,M$. Here $f_b$ 
 is the cosmic baryon fraction, and the efficiency $\epsilon(M)$ for the conversion of baryons into stars is taken from   \citealp{Mason2015}  (see their fig. 1). This is a redshift-independent relation characterized by a maximal efficiency at masses $M\approx 10^{12}\,M_{\odot}$, and constitutes  a phenomenological  representation of our knowledge about galaxy formation before the JWST era. 
 
 The star formation rate is related to the UV luminosity $L$  
 through the relation $\dot m_*/{\rm M_{\odot}\,yr^{-1}}=k_{UV}\,L/\,{\rm erg\,s^{-1}\,Hz^{-1}}$ with $k_{UV}=0.7\,10^{-28}$  \citep{Madau2014},  so that $L\propto \epsilon(M)\,M$. The galaxy luminosity functions are then computed as 
 $\phi (L)= dN(M)\,dM/dL$. 

 Adopting the above efficiency $\epsilon(M)$ to relate the DM mass and the UV luminosity results into a UV LF  that captures the evolution of the observed LF over all available observations for $0\leq z\leq 10$. It is also consistent with the LF observed by JWST for all redshifts $z\lesssim 9$ \citep[see, e.g.,][]{gelli}, while it under-predicts the abundances of bright galaxies measured by JWST at higher redshifts. Its behaviour is described by a 
Schechter form $\phi(L)=\phi_*\,(L/L_c(a))^{\alpha}\,exp(-L/L_c(a))$, where the normalization $\phi_*$, the logarithmic slope $\alpha$ and the evolution of the characteristic luminosity $L_c(a)$ are given in Mason et al. (2015) for the $\Lambda$CDM case. 
 When nCC cosmologies are assumed, the characteristic luminosity $L_c$ gets a boost over the value in the 
 $\Lambda$CDM case. A simple estimate of such a boost can be derived by noticing that  in the mass range $M\approx 10^9-10^{12}\,M_{\odot}$ relevant to the high-redshift galaxies considered here, the behaviour of $\epsilon (M)$ yields the approximate   relation $L/L_{\odot}\propto (M/M_{\odot})^{3/2}$ (see fig. 6 in  \citealp{Mason2015}). 
In this case, the luminosity function $\phi(L)$ in nCC can be derived from that given in  \citealp{Mason2015} simply by boosting the characteristic luminosity by a factor
$L_c/L_c(\Lambda CDM)\approx [M_c/M_c(\Lambda CDM)]^{3/2}$, where $M_c/M_c(\Lambda CDM)$ is the boost in the characteristic mass shown in fig. 1. In the redshift range $11\lesssim z\lesssim 15$ a boost in $M_c$ of a factor up $\approx 4$ thus results luminisities exceeding the 
$\Lambda$CDM expectations by factors up to $\approx 8$.
 
 Finally, we take into account that observed LFs are usually derived assuming volumes $V$ and luminosity distances $D_L$ inferred assuming $\Lambda$CDM model. Thus we convert   our luminosity functions to the value that they would have when interpreted by an observer that assumes a $\Lambda$CDM cosmology,   by multiplying the model LF and luminosities by a factor 
   $f_{Vol}=(dV/dz)/(dV_{nCC}/dz)$ and $f_{lum}=D^{2}_{L}/D^{2}_{nCC}$, respectively. 

\begin{figure*}[ht!]
\renewcommand{\thefigure}{3}
    \centering
    \includegraphics[width=0.6\linewidth,clip,angle=270]{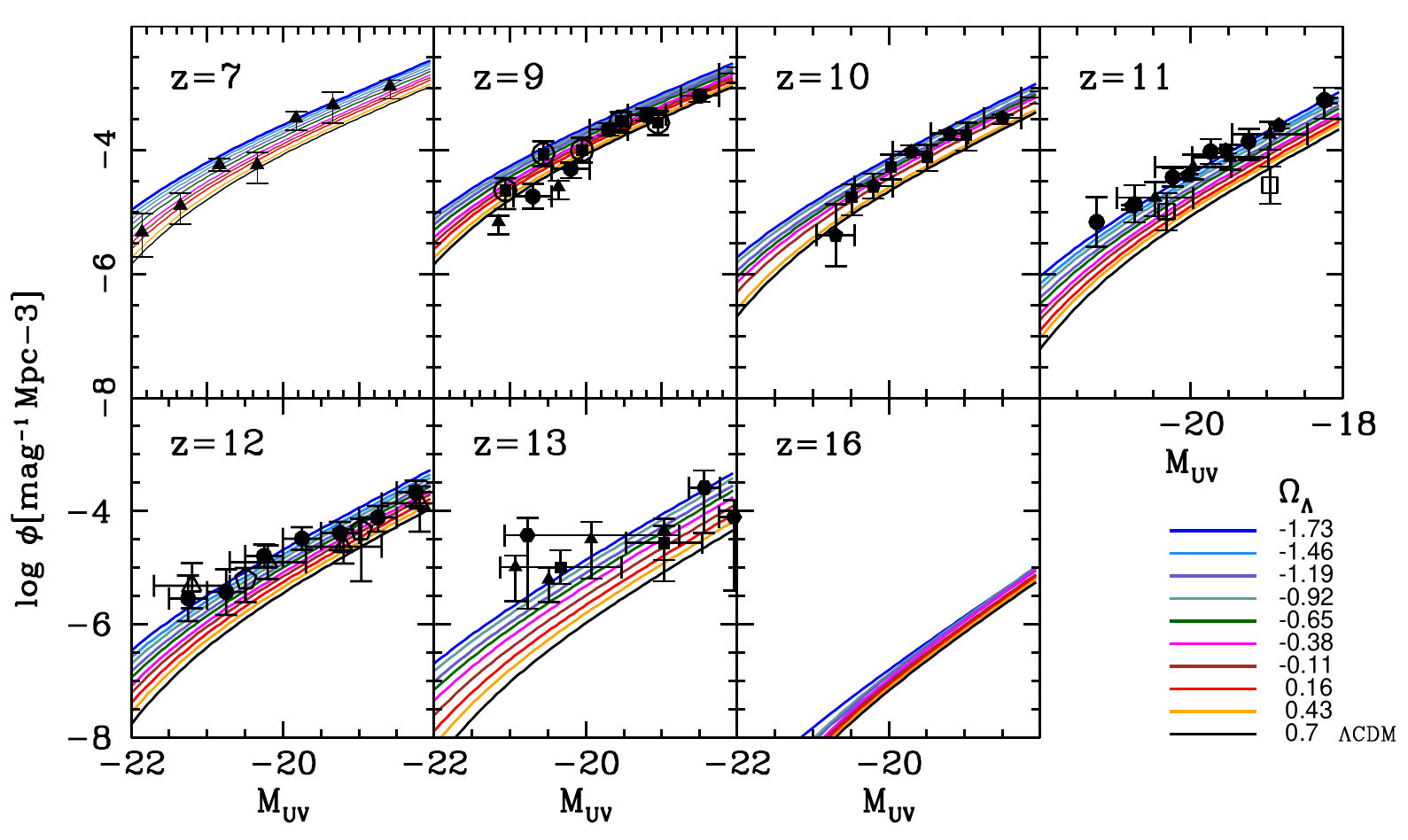}
    \caption{The evolution of galaxy LFs for the different values of $\Omega_{\Lambda}$ shown in the  legend. We assumed our fiducial choice of the combination $w_0=-0.9$, $w_a=0.1$. The data are from \citet{Bouwens2015,Bouwens2021,Bouwens2022}  (filled triangles), \citet{Finkelstein2023a} (filled squares), \citet{Stefanon2021}  (open circles), \citet{Donnan2023} (filled circles), \citet{McLeod2024} (filled pentagons),  \citet{Casey2024} (empty squares), \citet{Harikane2022b} (empty triangles), \citet{Adams2024} (empty pentagons); \citet{Robertson2023} (filled exagons).}
    \label{fig:fig3}
\end{figure*}
   
\section {Results}
In fig. 3 we show the resulting evolution of the LFs for cosmological models with different values of $\Omega_{\Lambda}$ and for our fiducial choice of the combination $(w_0,w_a$).
While for $z\lesssim 10$ the LFs obtained for different values of $\Omega_{\Lambda}$ do not show large differences compared to the  $\Lambda$CDM case (the black line), at larger redshifts the boost in the mass distribution shown in figs. 1, 2 brings the LFs in much better agreement with data, without changing the star formation prescription  with respect to that holding at lower redshifts. Notice that at redshifts larger than $z= 16$ the effect of assuming nCC cosmologies becomes negligible, and the luminosity functions of all nCC models become similar to that predicted in the $\Lambda$CDM case. This is a direct consequence of the behaviour of the characteristic mass shown in fig. 1, as explained in Sect. 3, and constitutes a clear prediction of 
our study that can be tested with future results from JWST.

The corresponding evolution of the luminosity density of the whole galaxy population is shown in Fig. 4, and compared with different data sets. While a detailed best-fit approach is beyond the demonstrative scope of this paper, we notice that the values of $\Omega_{\Lambda}$ which provide a good match to the observed LFs and to the luminosity density in all redshift bins  can be qualitatively estimated as $\Omega_{\Lambda}\approx(-0.6,-0.3)$. 
It is extremely interesting that such a range of values is close to the preferred range of $\Omega_{\Lambda}$  obtained from the analysis of the recent DESI data when $\Omega_{\Lambda}$ is allowed to vary \citep{Wang2024}. This is particularly noticeable, since in principle $\Omega_{\Lambda}$ can take any value. 

 Fig. 4 clearly enlightens the sharp prediction of nCC models 
discussed above, namely, the 
fast decline of the boost in mass and luminosity with respect to $\Lambda$CDM predictions for $z\gtrsim 15$, where 
 the evolution of the LFs and luminosity density of the Universe are expected to merge with that envisaged by $\Lambda$CDM. This behaviour constitutes a clear way to disentangle the cosmological effects considered here from the astrophysical processes which might also affect the evolution of the LFs at early epochs. 

Finally, we notice that our conclusions are not depending on our specific choice of a fiducial combination $(w_0,w_a)$. Indeed, we show in Fig. 5 the effect of varying the $(w_0,w_a)$ combination on the values of $\Omega_{\Lambda}$ which provide (within 5\% accuracy) the same boost in 
the maximum characteristic mass (and hence the same LFs) of our fiducial choice. 
It is seen that, within the constraints on ($w_0,w_a$) provided by current DESI+Planck+SNIa data, values $\Omega_{\Lambda}\approx (-0.6,-0.3)$ are obtained. Remarkably, this is consistent with that obtained from the analysis of DESI data in the overlapping region of the ($w_0,w_a$) plane \citep{Wang2024}.

\begin{figure}
\renewcommand{\thefigure}{4}
\vspace{-0.1 cm}
\hspace{-0.3cm}
\includegraphics[width=1.05\linewidth,height=7.5cm]{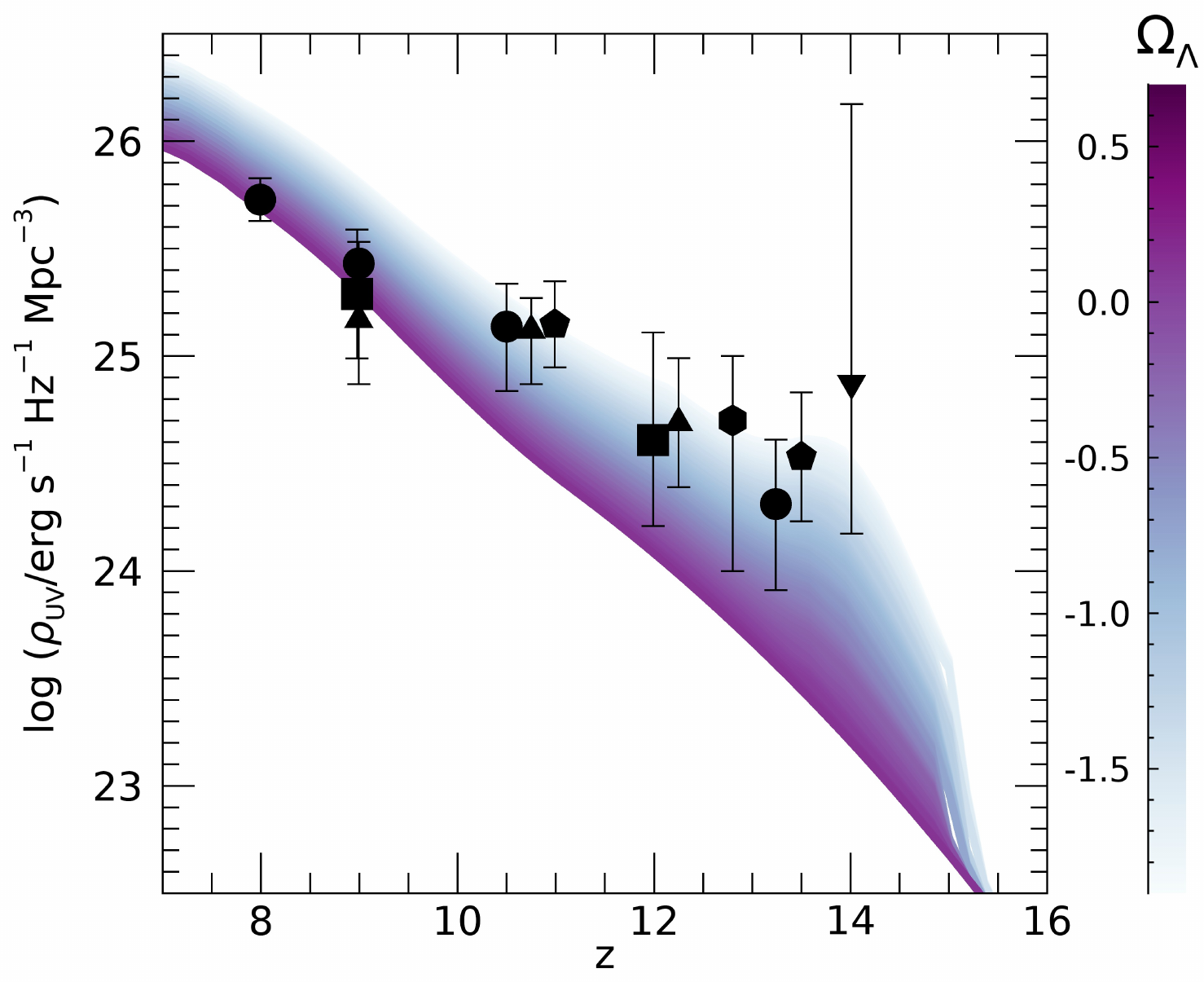}
\caption {The evolution of UV luminosity density for the values of $\Omega_{\Lambda}$ shown in the color bar, and for our fiducial combination $w_0=-0.9$, $w_a=0.1$. 
 The data are from \citet{Finkelstein2023a} (downward triangles),  \citet{Donnan2024} (circles), \citet{McLeod2024} (filled pentagons), \citet{Harikane2022b} (squares), \citet{Adams2024} (hexagons), \citet{PerezGonzalez2023} (upward triangles).}
 \label{fig:fig4}
 \end{figure}

\begin{figure}
\renewcommand{\thefigure}{5}
\vspace{-0.15 cm}
\hspace{-0.5cm}
\includegraphics[width=1.06\linewidth,height=7.8cm]{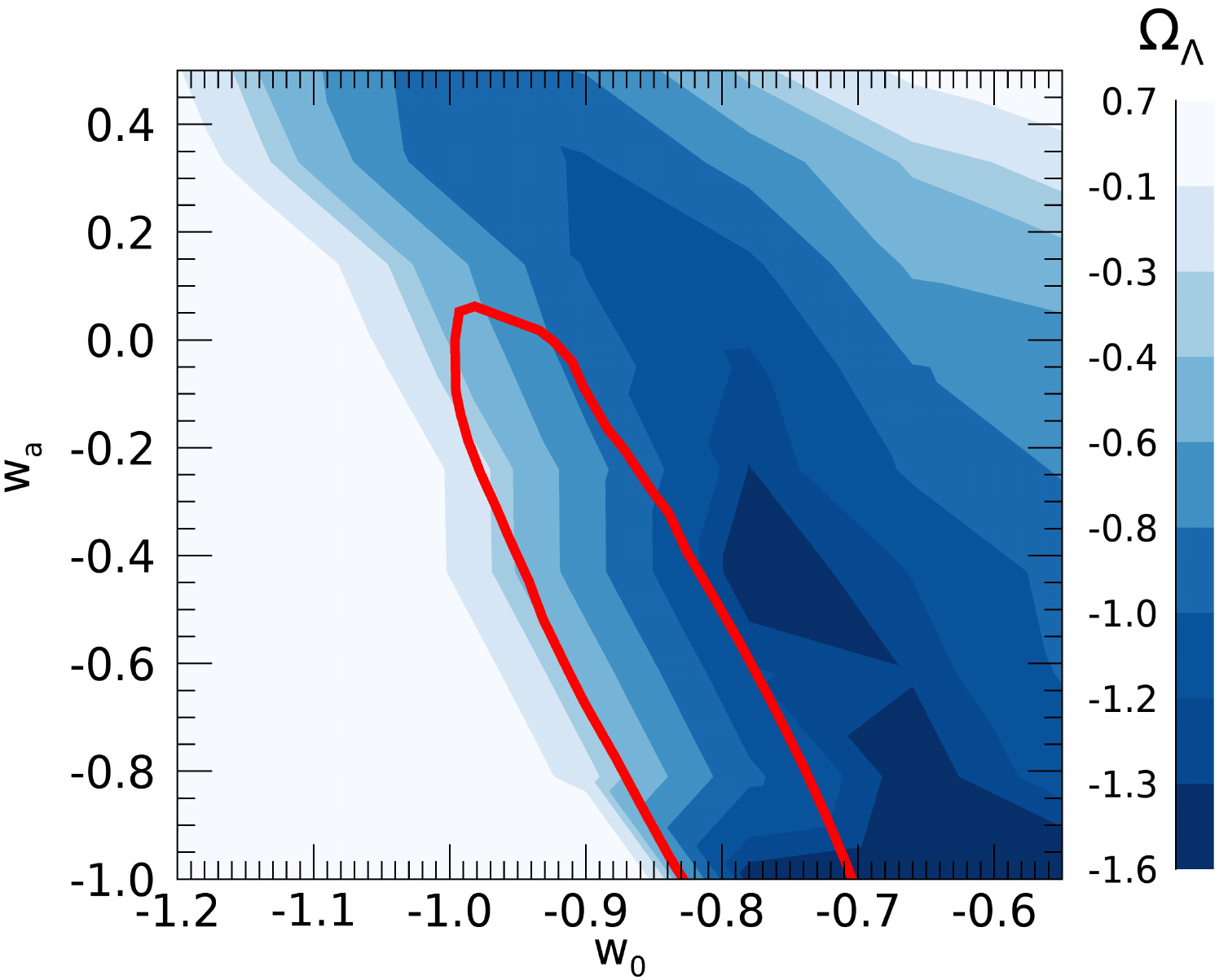}
\caption  {For each $(w_0,w_a)$ we show as coloured contours the values of $\Omega_{\Lambda}$ leading to the same boost (to within $5\%$) in 
the peak value of the characteristic mass shown in fig. 1. The red contour shows the 2-$\sigma$ confidence region consistent with  CMB+DESI+Pantheon Plus datasets \citep{Wang2024}.
}
\label{fig:fig5}
\end{figure}


\section{Discussion}
In this paper we propose that two ground-breaking recent observational results -  the indications for evolving DE resulting from combined cosmological probes, and the excess of bright sources at high redshifts $z\gtrsim 10$ compared to pre-JWST expectations - stem from a  unique cosmological origin, tracing back to a DE with negative $\Lambda$. Here we discuss the robustness of the two  
observational results above, and their different interpretations in the literature. 

The strong ($\sim 3.9\sigma$) evidence for DE is based on the combination BAO and Planck observations with the  Dark Energy 5 year (DES5Y) SN sample. While recent works have raised the possibility of an incorrect calibration for this sample \citep{efstathiou24}, tensions at the level of (or larger to) 
 $\sim 2.5\sigma$ persist when  different SNae samples (PantheonPlus, Union3) are considered.  While the assumption of CPL parametrisation may indeed affect the conclusions, it is interesting to note that 
 the region of the parameter space ($w_0$, $w_a$) favoured by the above combined cosmological probes is also favoured by the observations of massive galaxies at redshifts $z\approx 6-9$ \citep{Menci2024}, and that 2-$\sigma$ indications for dynamical DE also result when BAO compilations from the completed Sloan Digital Sky Survey (SDSS) are adopted as late- time cosmological observables \citep{mukhopa24}. In addition, independent analysis \citep[see, e.g.][]{raam2} prior to the last DESI results considered different time-changing $w(a)$ beyond the CPL approximation (in particular, linear potential models), also found dynamical DE solutions to be favoured compared to $\Lambda$CDM. 
  
 As for the recent estimates of the UV LF, the excess of bright sources compared to pre-JWST expectations has been shown on the basis of spectroscopic follow up to be robust against potential systematics in target selection and redshift uncertainties \citep[e.g.,][]{Harikane2024}, and it is unlikely explained by cosmic variance effects being found at high-significance in several, independent survey fields \citep[e.g.,][]{McLeod2024}. A possible explanation could reside in the contribution of Active Galactic Nuclei (AGN) to the UV emission of high redshift galaxies \citep[e.g.][]{hedge}, but this would require black holes  over-massive with respect to their host galaxies compared to the local relation (in the specific redshift range $z\gtrsim 10$).  Although  it has been suggested that this might indeed be the case at 
 high-redshifts \citep[see][]{Maiolino2023ref}, in current samples the majority of the detected objects have extended morphologies, suggesting that an AGN is not the dominant source of luminosity \citep[e.g.,][]{Harikane2024b}.
  
 On the other hand, several physical scenarios have been proposed potentially producing a slower evolution of the UV LF that either require peculiar, short-lived evolutionary phases or a sudden change in the underlying star-formation processes. It has been suggested that the high UV luminosities highlight an increase in the star-formation efficiency at high-redshift \citep[e.g.,][]{Finkelstein2023b} which in turn may be due to the occurrence of feedback-free starbursts, i.e. efficient star-formation on timescales shorter than the typical timescale to develop winds and supernovae \citep[][]{Dekel2023}. As an alternative, it has been suggested that the slow evolution of the UV LF beyond z$\sim$9 is explained by strong radiation-driven outflows in a short-lived, high sSFR "super-Eddington phase" which clears the objects from the previously formed dust \citep[][]{Ferrara2023,Ferrara2024,Fiore2023}. A similar boosting effect on the UV LF may result from an increased stochasticity of the star-formation histories at very high-redshift \citep[e.g.,][]{Mason2022,Kravtsov2024}. However, recent efforts to include such effects into cosmological semi-analytic models of galaxy formation \citep{yung24} have failed to account for the observed excess even assuming dust-free models. E.g., they showed that the inclusion of  stochastic bursts of star formation  would require a rather large stochastic component ($\sigma_{UV}\approx 2$, where $\sigma_{UV}$ is the root variance of a Gaussian random deviate in UV magnitude) to account for the observed excess. This is much larger than the stochasticity produced in the high-resolution radiation-hydrodynamic cosmological simulations of \cite{pallottini23}, which yield typical $\sigma_{UV}\approx 0.6$. Notice that such models do not include any suppression of star formation by the UV background at $z\gtrsim 8$, yet they still under-predict the observed counts. Thus the proposal that the observed excess could be explained by the lack of suppression of star formation via photo-ionization before reionization seems also to fail in providing a complete explanation of the observations. Other recent studies of galaxy formation in a cosmological context based on hydrodynamic simulation report 
  basically the same conclusions, see, e.g., \cite{kannan},\cite{wu2020}.

 Finally, more mundane explanations rely on an increased UV luminosity due to the presence of emission from PopIII stars or AGN \citep[][]{Harikane2022b}, or a top-heavy IMF \citep[][]{Trinca2024}. While all the above mentioned scenarios provide viable astrophysical explanations to the measured excess in the UV LF, they postulate a somewhat sudden change in the galaxy properties or physical processes in the first $\sim$500 Myr after the Big Bang. In the present work we have shown that the boosting effect on DM masses due to a negative $\Lambda$ yields to UV LFs that are compatible with recent estimates without requiring any modification in the underlying baryonic processes. Other  explanations that have been proposed that do not postulate a substantial change in galaxy formation processes propose a enahncement of the  power spectrum on scales of $\sim$1 Mpc \citep[][]{Padmanabhan2023}, a different time-redshift relation \citep[][]{Melia2023,Melia2024} or an accelerated formation of galaxies and clusters in MOND cosmologies \citep[][]{McGaugh2024}.  However, such alternatives at the moment are either based on ad-hoc modifications of some cosmological quantities, or lack a comprehensive theoretical framework of the underlying physical mechanisms. 
 Compared to the theoretical works mentioned above, the agreement between the range of values for $\Omega_{\Lambda}$ needed to match the observed LFs in nCC cosmologies discussed in the present paper and that obtained from independent cosmological probes provides a tantalizing perspective. In addition, the cosmological scenario we propose allows to {\it simultaneously}  account not only for the recent DESI results and for the observed abundance of UV luminous galaxies at $z\gtrsim 10$, but also for the unexpectedly large number of massive galaxies at $z\gtrsim 6$, an observational results which now being confirmed by spectroscopic data and which, although marginally consistent with $\Lambda$CDM predictions, appreciably favours phantom models \citep{Menci2022}, or models with nCC \citep{Menci2024}. 

In this context, disentangling between phantom  and nCC models is not an easy task. In fact, as noted in the Introduction, phantom models also simultaneously account for the same wide set of observations. 
On the one hand, on the theoretical side, phantom models are difficult to justify in terms of fundamental physics (see Sect. 1), while AdS vacua are ubiquitous features of holographic scenarios for gravity and string models. On the other hand, on the observational side,   
the sudden drop in the abundance of luminous galaxies shown in figs. 3 and 4 constitute a clear prediction of nCC models in the context of 
the CPL parametrization. However, 
\cite{tada2024} have shown that the $w_0-w_a$  
parameter space for CPL parametrisation as constrained by DESI observations can be mapped to a quintessence scalar field with a potential having a negative or AdS minimum. 
Thus a more rigorous approach toward disentangling between the two cosmological scenarios will necessarily require an analysis beyond the CPL parametrisation, and approach we plan to take on in forthcoming works. 

Future efforts aimed at discriminating among the above mentioned scenarios will need to be based on a multifaceted approach. The models postulating a change in galaxy evolution processes at z$\gtrsim$9 will need to be tested through a detailed measurement of their predictions on galaxy properties such as metallicity, specific star formation rate, gas conditions, dust obscuration, prevalence of AGN emission. On the other hand, a promising way to discriminate among various scenarios is to test predictions on the abundance of bright galaxies at earlier epochs.  Pushing the constraints on the UV LF at z$\gtrsim$ 15 is challenging, albeit within reach of JWST instruments \citep[][]{Conselice2024}. In this respect, as shown in the present work, galaxy evolution in nCC cosmologies presents the very clear prediction of a sharp decrease of galaxy abundance at redshifts higher than those probed so far.

\section{Conclusions}
Motivated by recent breakthroughs in cosmology resulting from 
combined cosmological probes, we have considered the effect of assuming cosmological models with a DE sector containing a negative $\Lambda$ as a ground state of the quintessence field on the galaxy luminosity function at high redshifts. Our main results can be summarized as follows. 
\newline
$\bullet$ The DM masses of galaxies in the redshift range $10\lesssim z\lesssim 15$ are boosted with respect to the $\Lambda$CDM expectations by a 
 factor $2-4$ depending of the value of $\Omega_{\Lambda}$. This  approximately corresponds to a boost in UV luminosity  $\approx M^{3/2}\approx 3-8$ (see discussion at the end of Sect. 3).
\newline
$\bullet$  When luminosities are related to the DM mass using standard relations that proved to match to the LFs at lower redshifts, the boost in the DM mass characterizing nCC models yields LFs which are able to match the LFs observed by JWST for $10\lesssim z\lesssim  15$,  without need to implement new physics to relate DM mass to the star formation. 
The sensitivity of the LFs to the value of $\Omega_{\Lambda}$ makes them a valuable tool to measure such a quantity.
\newline
$\bullet$ The range of values $\Omega_{\Lambda}=[-0.6,-0.3]$ needed to match the observed LFs at $10\lesssim z\lesssim  15$ in nCC cosmlogies agrees with that  obtained from the combined analysis of the recent DESI data with existing independent cosmological probes. 
\newline
$\bullet$ nCC models thus affect the UV luminosities in the right redshift range needed to match the JWST observations, and with the correct value of $\Omega_{\Lambda}$ which is required by independent cosmological probes \citep{Wang2024}. 
\newline
$\bullet$ A sharp prediction of nCC models for the evolution of the LFs and luminosity density of the Universe is the sharp decline of the boost in mass and luminosity with respect to $\Lambda$CDM predictions at epochs 
earlier than $z\approx 15$. This behaviour constitutes a clear way to disentangle the cosmological effects considered here from the astrophysical processes which might also affect the evolution of the 
LFs at early epochs. 

\begin{acknowledgements}
We acknowledge support from INAF Mini-grant ``Reionization and Fundamental Cosmology with High-Redshift Galaxies" and from the INAF Theory-grant "AGN-driven outflows in cosmological models of galaxy formation". AAS acknowledges the funding from SERB, Govt of India under the research grant no: CRG/2020/004347. AAS also acknowledges supports from INAF - Osservatorio Astronomico di Roma, Rome, Italy, CERN, Switzerland and Abdus Salam International Centre For Theoretical Physics, Trieste, Italy, for his visits where part of the work have been done. We thank the referee for useful comments that helped to improve the manuscript. 

\end{acknowledgements}

\bibliography{main}{}
\bibliographystyle{aasjournal}

\end{document}